\documentclass[%
 reprint,
 superscriptaddress,
 amsmath,amssymb,
 aps,UTF8,
]{revtex4-1}

\usepackage{graphicx}
\usepackage{dcolumn}
\usepackage{bm}
\usepackage{xcolor}
\usepackage{url}
\usepackage{float}
\usepackage{array}

\usepackage{booktabs}

\usepackage{subfigure}
\usepackage{multirow}

\begin{document}


\title{Extraction of Proton Trace Anomaly Energy from Near-Threshold $\phi$ and $J/\psi$ photo-productions}
\author{Wei Kou}
\affiliation{Institute of Modern Physics, Chinese Academy of Sciences, Lanzhou 730000, China}
\affiliation{University of Chinese Academy of Sciences, Beijing 100049, China}

\author{Rong Wang}
\email{rwang@impcas.ac.cn}
\affiliation{Institute of Modern Physics, Chinese Academy of Sciences, Lanzhou 730000, China}
\affiliation{University of Chinese Academy of Sciences, Beijing 100049, China}

\author{Xurong Chen}
\email{xchen@impcas.ac.cn}
\affiliation{Institute of Modern Physics, Chinese Academy of Sciences, Lanzhou 730000, China}
\affiliation{University of Chinese Academy of Sciences, Beijing 100049, China}
\affiliation{Guangdong Provincial Key Laboratory of Nuclear Science, Institute of Quantum Matter, South China Normal University, Guangzhou 510006, China}



\begin{abstract}
The trace anomalous energy contribution to the proton mass is 
a very important topic in non-perturbative QCD and hadron physics.
In experiments, it is under the hot discussions on 
how to measure the trace anomalous energy.
The QCD interpretation of proton trace anomaly is still not clear.
To connect the theory with the experiment,
we extract the trace anomaly by analyzing the near-threshold photo-production 
data of $\phi$ and J/$\psi$ vector mesons. 
Based on the vector-meson-dominance model and QCD Van der Waals representation, 
we find that the percentage of trace anomaly in the proton mass 
ranges from $16\%$ to $24\%$, which is of similar order of magnitude 
as the 23$\%$ given by Lattice QCD. 
We also provide the approximate magnitudes of the systematic uncertainties 
of the extracted results from the model assumptions as well as the data fitting procedures. 
We give relative statistical uncertainties of 17.2$\%$, 17.7$\%$, 3.6$\%$, and 8.2$\%$, 
total relative systematic uncertainties of 21.4$\%$, 54.2$\%$, 37.2$\%$, and 25.7$\%$, 
for the analyses of the GlueX, LEPS, CLAS, and SAPHIR data, respectively. 
We argue that the near-threshold $\Upsilon$ photo-production experiments are 
more beneficial for the measurement of trace anomaly in proton mass in the future.
\end{abstract}

\maketitle


\section{Introduction}
\label{sec:intro}

The majority of the visible mass of the universe resides 
in the two types of nucleons -- protons and neutrons.
Nucleons are made of massless gluons and almost massless quarks.
The generation of nucleon mass is one of the puzzles in modern particle physics.
The origin of the masses of fundamental particles (leptons, quarks, and massive gauge bosons)
are delicately explained by the famous Higgs mechanism \cite{Higgs:1964pj,Guralnik:1964eu,Englert:1964et}.
The proton mass comes mainly from the complicated workings
of non-perturbative Quantum Chromodynamics (QCD) \cite{Gross:1973id,Gross:1973ju,Gross:1974cs}.
QCD theory originates from Yang-Mills theory \cite{Yang:1954ek} in the 1950s,
which belongs to a non-abelian gauge theory.
To calculate the proton mass in the first principle view,
one encounters two main challenges:
the color confinement for the constitutes (quarks and gluons) \cite{Wilson:1974sk}
and the divergent strong coupling constant at low energy scale.

The proton is a composite particle made of quarks and gluons,
hence its mass is usually argued to be from some different sources. 
The concept of trace anomaly was first introduced from Refs. \cite{Peskin:1995ev,Kharzeev:1995ij}. 
In Refs. \cite{Ji:1994av,Ji:1995sv}, Ji first define the proton mass decomposition
with QCD Hamiltonian operators.
Four gauge-invariant Hamiltonian operators are introduced as,
	\begin{equation}
		H_{\mathrm{QCD}}=H_{q}+H_{g}+H_{m}+H_{a},
		\label{eq:HamiltionQCD}
	\end{equation}
where
	\begin{equation}
		\begin{aligned}
			H_{q} &=\int d^{3} \vec{x}\left[\psi^{\dagger}(-i \mathbf{D} \cdot \alpha) \psi\right], \\
			H_{g} &=\int d^{3} \vec{x} \frac{1}{2}\left(\mathbf{E}^{2}+\mathbf{B}^{2}\right), \\
			H_{m} &=\int d^{3} \vec{x} \left(1+\frac{1}{4}\gamma_{m}\right) \bar{\psi} m \psi, \\
			H_{a} &=\int d^{3} \vec{x} \frac{1}{4} \beta(g)\left(\mathbf{E}^{2}-\mathbf{B}^{2}\right).
		\end{aligned}
	\label{eq:Hamiltoniandecom}
	\end{equation}
In Ref. \cite{Ji:1995sv}, the author assumed that the hadron
mass is calculated as the expectation value of the Hamiltonian
at the hadron rest frame:
	\begin{equation}
		M_N=\left.\frac{\left\langle P\left|H_{\mathrm{QCD}}\right| P\right\rangle}{\langle P \mid P\rangle}\right|_{\text {rest frame }},
		\label{eq:Hardonstate}
	\end{equation}
which is decomposed into four terms characterized by
the QCD trace anomaly parameter $b(\mu^2)$
and the momentum fraction $a(\mu^2)$ carried by all quarks \cite{Ji:1995sv}.
The four terms of the proton mass partitions are written as \cite{Ji:1995sv},
	\begin{equation}
		\begin{aligned}
			&M_{q} =\frac{3}{4}\left(a-\frac{b}{1+\gamma_{m}}\right) M_{N}, \ \
			M_{g} =\frac{3}{4}(1-a) M_{N}, \\
			&M_{m} =\frac{4+\gamma_{m}}{4\left(1+\gamma_{m}\right)} b M_{N}, \ \
			M_{a} =\frac{1}{4}(1-b) M_{N},
		\end{aligned}
		\label{eq:Protom_mass}
	\end{equation}
where the anomalous dimension of quark mass $\gamma_{m}$ \cite{Buras:1979yt} describes the renormalization information.
The momentum fraction $a(\mu^2)$ of the quarks is easily computed 
with all the quark distributions determined in experiments as,
	\begin{equation}
		a\left(\mu^{2}\right)=\sum_{f} \int_{0}^{1} x\left[q_{f}\left(x, \mu^{2}\right)+\bar{q}_{f}\left(x, \mu^{2}\right)\right] d x.
		\label{eq:Momentum_frac}
	\end{equation}

The proton mass decomposition is from the analysis of 
the energy-momentum tensor (EMT) in QCD theory \cite{Ji:1995sv}.
The first three terms in Eq. (\ref{eq:Protom_mass}) can be easily 
understood with the classical field theory.
However the last term is an extension of the classical description 
in the quantum field theory -- the quantum anomaly.
In the recent papers \cite{Ji:2021mtz,Ji:2021pys}, the authors provide further discussion
on the the source of quantum anomalous energy (QAE) to the proton mass 
and argue that it arises from the scale symmetry breaking due to 
the ultraviolet (UV) divergences in the quantum field theories.
The trace anomaly part to the proton mass resembles 
the dynamical Higgs mechanism to the mass of the fundamental fermions.
The author mentions that the quantum anomalous contribution is 
scale-independent in the chiral limit \cite{Ji:2021mtz}.
However the quark mass is not exactly zero.
Therefore, the quantum anomalous energy should be scale-dependent due to
the contribution of the quark mass \cite{NOVIKOV198067,Novikov:1981xi}.

Although the proton mass is mainly generated by the dynamical chiral symmetry breaking,
a small portion does arise from the masses of the quarks.
The quark mass contribution is usually characterized by a matrix element parameter $b$ \cite{Ji:1995sv}, 
which is used to represent the Hamiltonian of quark mass term 
in Eqs. (\ref{eq:Hamiltoniandecom}) and (\ref{eq:Protom_mass}). 
The parameter $b$ is related to the quark masses themselves
and the quark scalar charges of the proton \cite{Ji:1995sv,Kharzeev:1995ij},
	\begin{equation}
		\begin{aligned}
			b M_{N} &=\left\langle N\left|m_{u} \bar{u} u+m_{d} \bar{d} d\right| N\right\rangle+\left\langle N\left|m_{s} \bar{s} s\right| N\right\rangle \\
			&=m_{l}\langle N|\bar{u} u+\bar{d} d| N\rangle+m_{s}\langle N|\bar{s} s| N\rangle \\
			&=\sigma_{\pi N}+\sigma_{s N}.
		\end{aligned}
	\label{eq:sigmasn}
	\end{equation}
The scalar nucleon matrix element of up and down quarks $\sigma_{\pi N}$ is about 45 MeV,
determined by the low energy $\pi-N$ scattering amplitude \cite{Gasser:1990ce,Hoferichter:2015dsa,Ling:2017jyz}.
In Ref. \cite{Alarcon:2011zs}, the authors present a new result using new experimental data
and chiral effective field theory $\sigma_{\pi N}=59(7)$ MeV. 
Phenomenological estimation using $\pi-N$ scattering data was performed in the Ref. \cite{Pavan:2001wz}.
Too little is known about the $\sigma_{s N}$, which describes the strange
scalar charge $m_{s}\langle N|\bar{s} s| N\rangle$ in the nucleon \cite{Ji:1994av,Kharzeev:1995ij}.
It is believed to be large because the strange quark is heavier. 
However, in theory, the QCDSF Collaboration finds $\sigma_{s N}=11\pm13$ MeV \cite{Bali:2011ks}
and the $\chi$QCD Collaboration finds $\sigma_{s N}=40\pm12$ MeV \cite{Yang:2015uis}.
The strange part of nucleon matrix element $\sigma_{s N}$ is also suggested to be 
small around 16 MeV from an effective field theory \cite{Alarcon:2012nr}.
The nucleon scalar charge is an important parameter for calculating
the scattering between nucleon and the dark matter particles
\cite{Bottino:1999ei,PhysRevD.71.095007,PhysRevD.77.065026,PhysRevD.78.083520,Hill:2011be,PhysRevD.88.055025}.
Hence studying the QCD trace anomaly parameter $b$ might be helpful
in investigating the method of detecting the weakly-interacting dark matter particles.

At present, the direct experimental measurement of quantum effects 
inside protons is impossible to be realized.
But the low-energy scattering between heavy quarkonium and nucleon
can be used to probe the trace anomaly of the nucleon,
since the two-gluon exchange is dominant for the process 
\cite{NOVIKOV198067,Novikov:1981xi,Frankfurt:2002ka,Kharzeev:2021qkd}.
It is not difficult to compute the scattering amplitude using
the local operator product expansion (OPE) with the gluon operator \cite{Boussarie:2020vmu}.
Recently, in Refs. \cite{Hatta:2018ina,Hatta:2019lxo} the authors had tried to extract 
the trace anomaly from the J/$\psi$ photo-production data near threshold,
based on a holographic QCD framework.
Based on the Lattice QCD, the result of proton anomaly is discussed in Ref. \cite{PhysRevLett.121.212001}. 
For more information on the importance of the proton trace anomaly, 
one can refer to the EICUG Yellow Report \cite{AbdulKhalek:2021gbh}.
In this work, we begin with the vector-meson-dominance model (VMD) 
suggested by Refs. \cite{Kharzeev:1995ij,Kharzeev:1998bz}.
Following the recent analysis based on this method \cite{Wang:2019mza},
we try to extract the trace anomalous energy of the proton firstly 
from the diffractive production data of both $\phi$ and J/$\psi$ vector 
mesons near the thresholds \cite{Mibe:2005er,Ali:2019lzf}. 
The paper is organized as follows. 
We describe the forward meson-nucleon scattering amplitude 
in the VMD model in Sec. \ref{sec:method}, relating the parameter $b$ to the photo-production data. 
We subsequently show our extraction of the trace anomaly part 
by fitting the experimental data in Sec. \ref{sec:results}. 
We generally discuss and analyze the uncertainties 
introduced by the model and data fitting in Sec. \ref{sec:error}. 
Finally we give the conclusions and discussions in Sec. \ref{sec:summary}.

\section{Method}
\label{sec:method}

With the VMD model \cite{SAKURAI19601}, the forward cross section
of the vector meson $X$ ($\phi$, J/$\psi$, $\Upsilon$, etc.)
photo-production on the nucleon target is formulated as \cite{Kharzeev:1998bz},
	\begin{equation}
		\begin{aligned}
			&\left.\frac{d \sigma_{\gamma N \rightarrow X N}}{d t}\right|_{t=0} \\
			&=\left.\frac{3 \Gamma\left(X \rightarrow e^{+} e^{-}\right)}{\alpha m_{X}}\left(\frac{k_{X N}}{k_{\gamma N}}\right)^{2} 
               \frac{d \sigma_{X N \rightarrow X N}}{d t}\right|_{t=0},
		\end{aligned}
	\label{eq:xsection}
	\end{equation}
where $\alpha=1/137$ denotes the fine structure constant,
$k_{a b}^{2}=\left[s-\left(m_{a}+m_{b}\right)^{2}\right]\left[s-\left(m_{a}-m_{b}\right)^{2}\right] / 4 s$
describes the center-of-mass momentum square of the corresponding two-body system,
and $\Gamma$ is the partial decay width of the $X$ meson decaying into a $e^+e^-$ pair.
The $\left.\frac{d \sigma_{X N \rightarrow X N}}{d t}\right|_{t=0}$ term
of elastic scattering in the forward limit is given by,
	\begin{equation}
		\left.\frac{d \sigma_{X N \rightarrow X N}}{d t}\right|_{t=0}=\frac{1}{64 \pi} \frac{1}{m_{X}^{2}\left(\lambda^{2}-m_{N}^{2}\right)}\left|F_{X N}\right|^{2},
	\end{equation}
where $\lambda=(p_Np_X/m_X)$ is the nucleon energy at the quarkonium rest frame \cite{Kharzeev:1998bz}.
$F_{X N}$ denotes the invariant amplitude of $X-N$ elastic scattering.
Using these definitions, the amplitude takes the form \cite{Kharzeev:1995ij,Kharzeev:1998bz},
	\begin{equation}
		\begin{aligned}
			F_{X N} & \simeq r_{0}^{3} d_{2}  \frac{8 \pi^{2}M_Nm_X}{27}\left(M_{N}-
                             \left\langle N\left|\sum_{i=u, d, s} m_{i} \bar{q}_{i} q_{i}\right| N\right\rangle\right) \\
			& = r_0^3d_2\frac{8\pi^2}{27}(1-b)M_N^2m_X,
		\end{aligned}
		\label{eq:amplitude}
	\end{equation}
which is dominated by the QCD trace anomaly part.
For low-energy scattering and in the chiral limit,
the mass of a hadron state comes purely from the quantum fluctuations of the gluons.
Away from the chiral limit, the factor $(1-b)$ is used to characterize
the QCD trace anomaly contribution to the proton mass.
In Eq. (\ref{eq:amplitude}), the ``Bohr" radius $r_0$ of 
the meson $X$ is given by \cite{Kharzeev:1995ij},
	\begin{equation}
		r_{0}=\left(\frac{4}{3 \alpha_{s}}\right) \frac{1}{m_{q}},
		\label{eq:bohr}
	\end{equation}
where $m_q$ represents the mass of quark (strange quark $s$ for the $\phi$ meson 
and charm quark $c$ for the J/$\psi$ meson). 
In this work, we choose the constituent quark mass (low-energy scale) 
for the extraction, e.g. $m_c = 1.67$ GeV and $m_s = 0.486$ GeV \cite{griffiths2008introduction}. 
We discuss the effect of quark mass selection on the results in the final section.
In Eq. (\ref{eq:amplitude}), the Wilson coefficient $d_2$ is found 
in Refs. \cite{Kharzeev:1995ij,Peskin:1979va,Kharzeev:1996tw} as,
	\begin{equation}
		d_{n}^{(1 S)}=\left(\frac{32}{N_{c}}\right)^{2} \sqrt{\pi} \frac{\Gamma\left(n+\frac{5}{2}\right)}{\Gamma(n+5)},
		\label{eq:wilson}
	\end{equation}
where $N_c$ is the number of colors.
The strong coupling constant $\alpha_{s}$ depends on the renormalization scale $\mu^2$,
and the scale is chosen to be the ``Rydberg" energy square $\epsilon_0^2$
for the bound state of the quarkonium $X$ \cite{Kharzeev:1995ij,Kharzeev:1996tw}.

The running strong coupling constant $\alpha_{\mathrm{s}}$ is an important
parameter in QCD evolution equations. In this work, we use a
renormalization-group-invariant process-independent effective strong coupling 
constrained by the calculation of LQCD \cite{Cui:2019dwv}. 
The effective strong coupling shows a saturated plateau
approaching the infrared region, which agrees well with the
the Bjorken sum rule with meson PDFs at low $\mu^2 (<$ 1 GeV$^2$) \cite{Cui:2019dwv}. 
The saturated effective strong coupling is given by \cite{Cui:2019dwv,Binosi:2016nme},
	\begin{equation}
		\alpha_{\mathrm{s}}\left(\mu^{2}\right) = \frac{4 \pi}{\beta_{0} \ln\left[\left(m_{\alpha}^{2}+\mu^{2}\right) 
                                                                        / \Lambda_{\mathrm{QCD}}^{2}\right]},
		\label{eq:coupling}
	\end{equation}
where $\beta_{0}=\left(33-2 n_{\mathrm{f}}\right) / 3$ refers to 
the one-loop $\beta$ function coefficient, $n_\mathrm{f}$ is the number of flavors, 
$m_{\alpha}=0.43$ GeV is the effective gluon mass owning to 
the dynamical breaking of scale invariance \cite{Cui:2019dwv}. 
For the QCD cutoff, we chose $\Lambda_{\mathrm{QCD}}= 0.34$ GeV \cite{Han:2020vjp}.
Based on this analysis, we use the saturated form of strong coupling 
in order to take into account of the non-perturbative effect.

Applying the theoretical and phenomenological framework discussed above,
we extract the QCD trace anomaly $M_a$ from the extrapolated value 
of the differential photo-production cross section at Mandelstam variable $-t=0$ GeV$^2$.
We take the differential cross section data
from the diffractive $\phi$ photo-production near the threshold 
published by the LEPS, SAPHIR and CLAS Collaborations \cite{Mibe:2005er,Barth:2003bq,CLAS:2013jlg,Dey:2014tfa}
and the diffractive J/$\psi$ photo-production data
near the threshold by GlueX Collaboration at Jefferson Laboratory (JLab) \cite{Ali:2019lzf}.
The experimental data are fitted with an exponential function
  \begin{equation}
    \frac{d\sigma}{dt} =\frac{d\sigma}{dt}|_{t=0}\times e^{-kt},
  \label{eq:exp}
\end{equation}
where $d\sigma/dt|_{t=0}$ represents the forward differential cross-section 
and $k$ denotes the slope parameter. 
Eq. (\ref{eq:exp}) describes the $t$-dependence of the cross section. 
The details of our studies and the results are shown in the next section, 
focusing on the formulae mentioned in this section.

\section{Extraction of trace anomaly}
\label{sec:results}

\begin{figure}[htbp]
			\centering
			\includegraphics[width=0.45\textwidth]{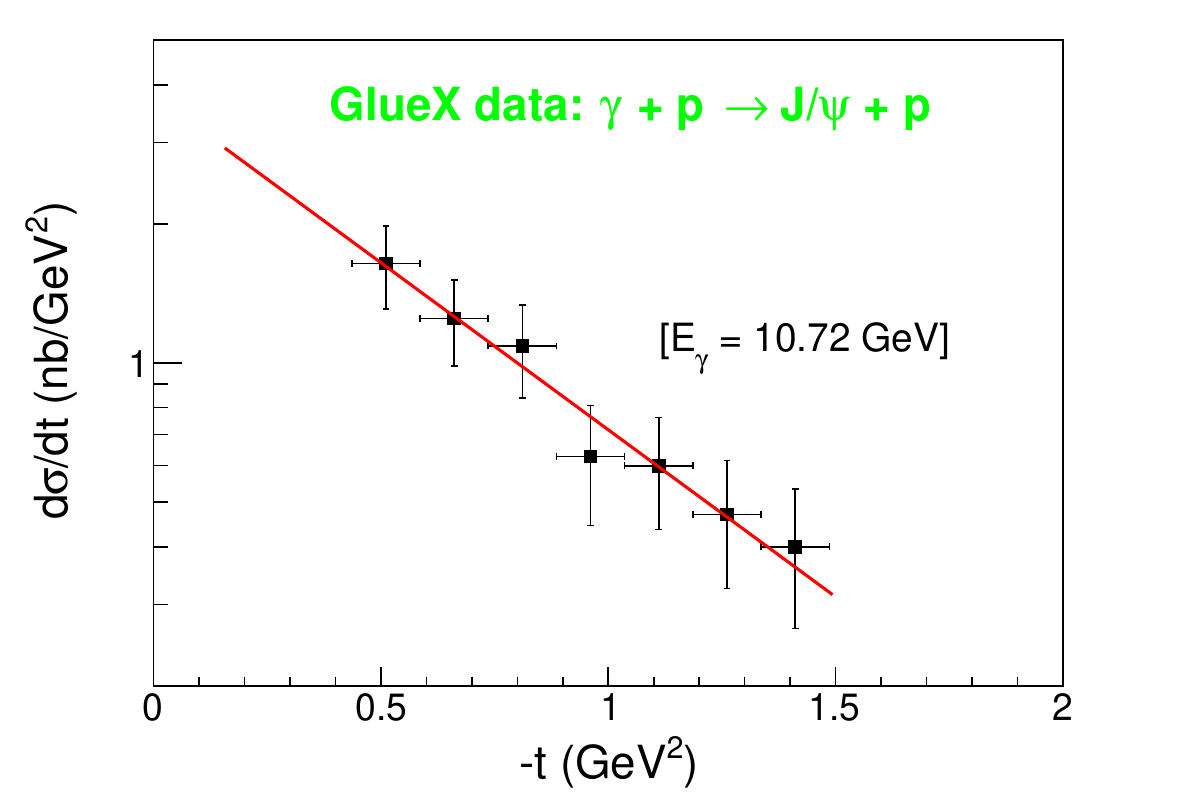}
			\caption{
				The differential cross section of J/$\psi$ photo-production near threshold
				of $E_{\gamma}=$ 10.72 GeV by GlueX Collaboration \cite{Ali:2019lzf}.
	            The red line shows our  fitting result with an exponential form. $\chi^2/d.o.f.=0.14$. Only statistical
	            uncertainties are shown. The obtained parameters are listed in TABLE \ref{tab:Gluex}.
			}
			\label{fig:Jpsi-GlueX}
		\end{figure}

Based on Eq. (\ref{eq:exp}), the forward differential cross-sections are 
obtained by fitting the experimental data of $\phi$ and J/$\psi$ photo-productions 
near the thresholds from Refs. \cite{Mibe:2005er,Barth:2003bq,CLAS:2013jlg,Ali:2019lzf}. 
According to Eqs. (\ref{eq:xsection}-\ref{eq:coupling}), 
the trace anomaly parts of proton mass are extracted by simple algebraic operations. 
FIG. \ref{fig:Jpsi-GlueX} shows the GlueX collaboration's near-threshold differential 
cross section data for J/$\psi$ photo-production on the proton. 
It is suggested that the $-t$-dependence of the differential cross section is 
well described with the exponential form \cite{Ali:2019lzf}.

TABLE \ref{tab:Gluex} lists the key parameters and the experimental settings
for the extraction of the QCD trace anomaly.
In this work we assume that the energy scale of the strong interaction ($\mu$) is 
equal to the ``Rydberg'' energy $\epsilon_0$ of the quark-antiquark pair \cite{Kharzeev:1996tw,Kharzeev:1998bz}, 
for the production near the threshold. 
For J/$\psi$ production, the binding energy $\epsilon_0^2=\mu^2=0.41$ GeV$^2$ is taken in Ref. \cite{Kharzeev:1995ij}. 
Thinking about pulling apart a $c\bar{c}$ pair to generate a $D\bar{D}$ pair, 
a naive estimate of the ``Rydberg" energy $\epsilon_0$ is $m_D +m_{\bar{D}}-m_{J/\psi}$ \cite{Kharzeev:1995ij,Wang:2019mza}. 
Similarly, we choose $\epsilon_0^2 = 0.14$ GeV$^2$ as the energy scale corresponding to the $\phi$ photo-production.

	\begin{table}[htbp]
		\caption{Forward cross section $\frac{d\sigma}{dt}|_{-t=0}$ and trace anomaly $M_a/M_N$ extracted
                 by fitting the experimental data from GlueX collaboration (corresponding to Fig. \ref{fig:Jpsi-GlueX}).
                 Only statistical uncertainties are considered.
                 The systematic uncertainty analysis is described in Sec. \ref{sec:error}.}
		\begin{center}
			\begin{tabular}{ |c|c|c|c|}
			\toprule[1.3pt]
			$E_\gamma$ (GeV) & $\frac{d\sigma}{dt}|_{-t=0}$ (nb/GeV$^2$)& $\chi^2/d.o.f.$&$M_a/M_N$ $(\%)$ \\
			\hline
			10.72& $3.79\pm1.32$&0.14&19.2$\pm$3.3\\
			\bottomrule[1.3pt]
		\end{tabular}	
		\label{tab:Gluex}
		\end{center}
	\end{table}

\begin{figure}[htbp]
	\centering
	\includegraphics[width=0.45\textwidth]{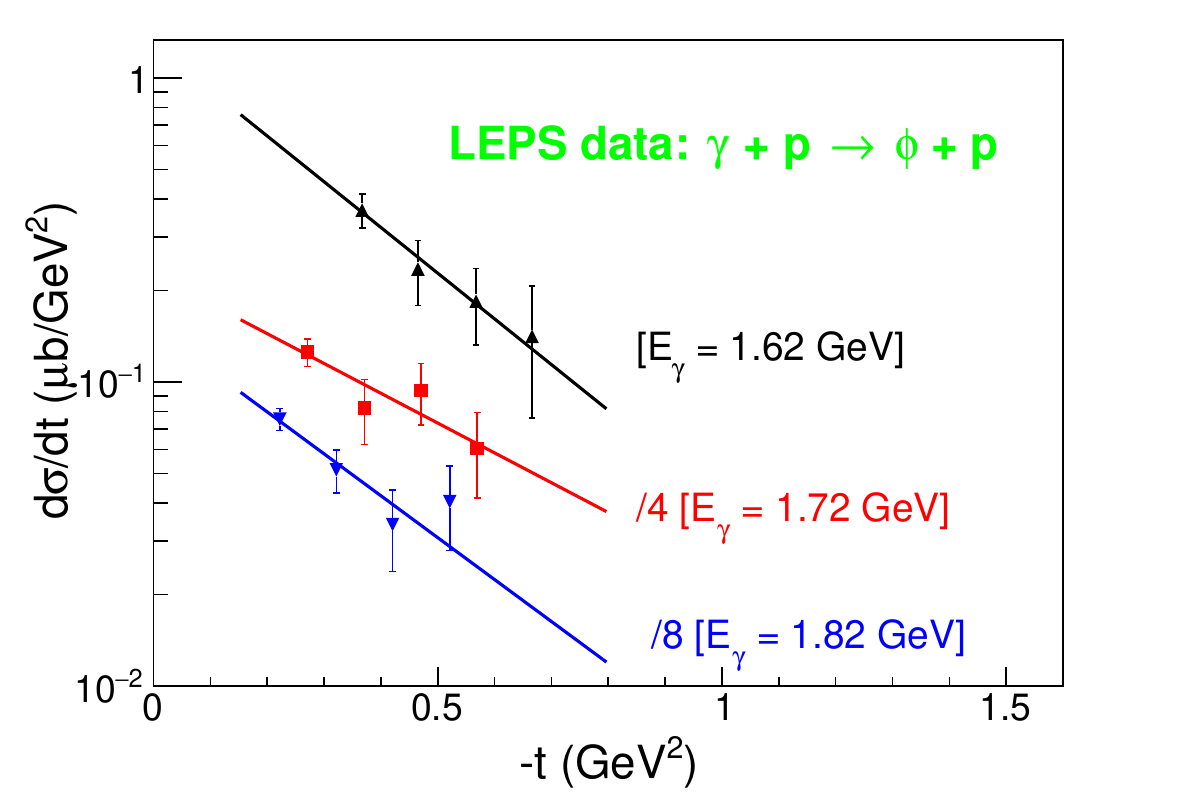}
	\caption{
		The differential cross sections of coherent $\phi$ photo-production on proton 
        near threshold ($\gamma p\to \phi p$) from LEPS \cite{Mibe:2005er}. 
        Only the statistical errors are presented. The cross-section
		data at $E_\gamma = 1.72$ GeV and $E_\gamma=1.82$ GeV are divided by 4
		and 8 respectively, as indicated in the figure. The solid lines show
		the fits of exponential form.
	}
	\label{fig:phi-LEPS}
\end{figure}

\begin{figure}[htbp]
	\centering
	\includegraphics[width=0.45\textwidth]{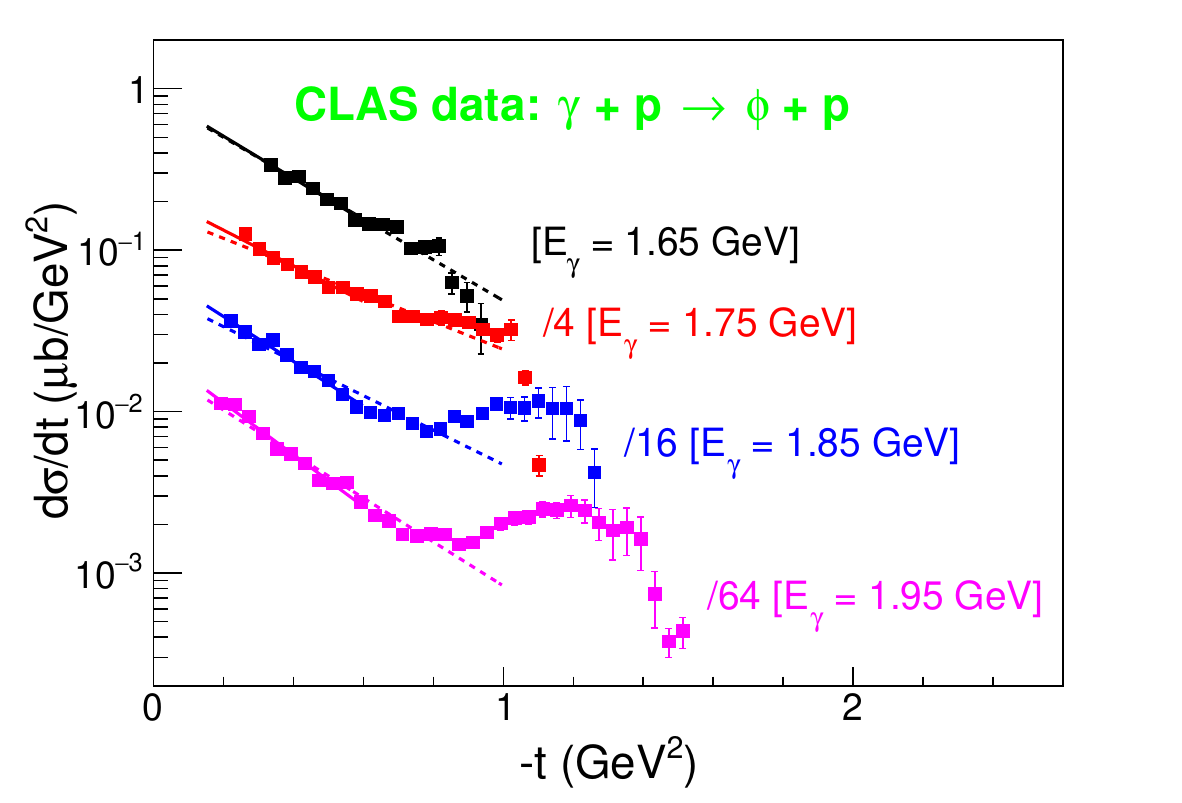}
	\caption{
		The differential cross sections of near-threshold $\phi$
		photo-production on a hydrogen target ($\gamma p\to \phi p$) from CLAS
		\cite{CLAS:2013jlg,Dey:2014tfa}. Only the statistical errors are presented. Some of the
		cross sections are scaled by factors indicated in the figure.
		The dashed lines show the fits in the $-t$ range from
		0.15 GeV$^2$ to 1 GeV$^2$. The solid lines show the fits in the
		$-t$ range from 0.15 to 0.6 GeV$^2$.
	}
	\label{fig:phi-Clas}
\end{figure}

The $\phi$ production data from CLAS, SAPHIR and LEPS collaborations are 
presented in FIGs. \ref{fig:phi-LEPS} and \ref{fig:phi-Clas}. 
We still use the exponential function fittings 
to get the forward differential cross sections, respectively. 
In the following paragraphs we present some of the details.

FIG. \ref{fig:phi-LEPS} shows the LEPS collaboration's near-threshold
differential cross section data for $\phi$ photo-production on the proton target.

The results for some important physical quantities are listed in TABLE \ref{tab:leps}.

\begin{table}[htbp]
	\caption{Forward cross section $\frac{d\sigma}{dt}|_{-t=0}$ and trace anomaly $M_a/M_N$ extracted
             by fitting the experimental data from LEPS collaboration (corresponding to Fig. \ref{fig:phi-LEPS}).
             Only statistical uncertainties are considered.
             The systematic uncertainty analysis is described in Sec. \ref{sec:error}.}
	\begin{center}
		\begin{tabular}{ |c|c|c|c|}
			\toprule[1.3pt]
			$E_\gamma$ (GeV) & $\frac{d\sigma}{dt}|_{-t=0}$ ($\mu$b/GeV$^2$)& $\chi^2/d.o.f.$&$M_a/M_N$ $(\%)$ \\
			\hline
			1.62& $1.29\pm0.68$&0.10&18.9$\pm$4.5\\
			\hline
			 1.72& $0.91\pm0.28$&0.60 & 16.5$\pm$2.5 \\
			 \hline
			1.82 &$1.20\pm0.33$ & 0.66& 20.0$\pm$2.8 \\
			\bottomrule[1.3pt]
		\end{tabular}	
		\label{tab:leps}
	\end{center}
\end{table}

\begin{figure}[htbp]
	\centering
	\includegraphics[width=0.45\textwidth]{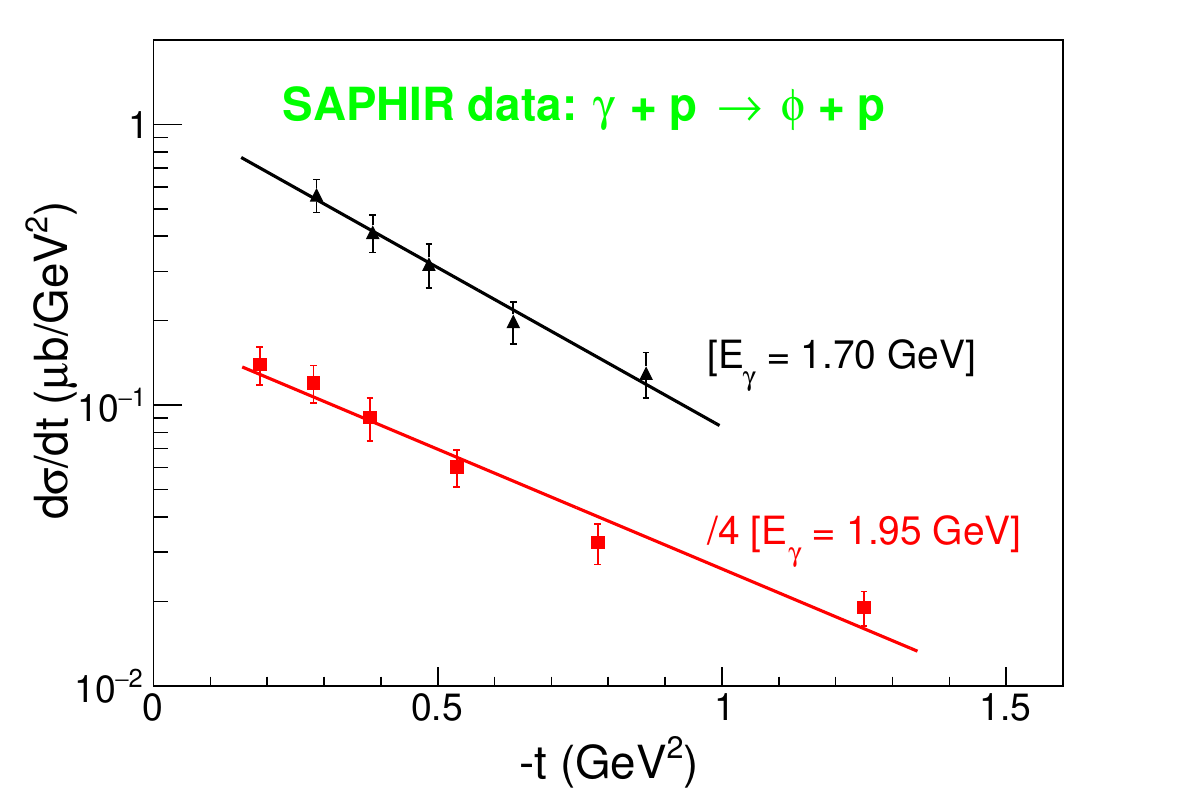}
	\caption{
		 The differential cross sections of coherent $\phi$ photo-production on proton 
         near threshold ($\gamma p\to \phi p$) from SAPHIR \cite{Barth:2003bq}. 
         Only the statistical errors are presented. The cross-section
         data at $E_\gamma = 1.95$ GeV is divided by 4, as indicated in the figure. 
         The solid lines show the fits of exponential form.
	}
	\label{fig:phi-saphir}
\end{figure}

We present the trace anomaly extracted from cross-section data 
at three energy points near the threshold measured by the LEPS collaboration \cite{Mibe:2005er}. 
Due to the low statistics, the uncertainty is relatively large.  
We also analyze the measurements from the SAPHIR collaboration \cite{Barth:2003bq} 
several decades ago for the comparisons. 
The differential cross section data of SAPHIR are shown in FIG. \ref{fig:phi-saphir}. 
From the fittings of the exponential form of Eq. (\ref{eq:exp}), 
the SAPHIR data give the proton trace anomaly parts to be $18.3\pm1.8\%$ and $16.9\pm1.1\%$ 
at $E_\gamma=1.7$ GeV and $E_\gamma=1.95$ GeV, respectively (summarized in TABLE \ref{tab:saphir}).

\begin{table}[htbp]
	\caption{Forward cross section $\frac{d\sigma}{dt}|_{-t=0}$ and trace anomaly $M_a/M_N$ extracted
             by fitting the experimental data from SAPHIR collaboration (corresponding to Fig. \ref{fig:phi-saphir}).
             Only statistical uncertainties are considered.
             The systematic uncertainty analysis is described in Sec. \ref{sec:error}.}
	\begin{center}
		\begin{tabular}{ |c|c|c|c|}
			\toprule[1.3pt]
			$E_\gamma$ (GeV) & $\frac{d\sigma}{dt}|_{-t=0}$ ($\mu$b/GeV$^2$)& $\chi^2/d.o.f.$&$M_a/M_N$ $(\%)$ \\
			\hline
			1.70& $1.14\pm0.23$&0.21&$18.3\pm1.8$\\
			\hline
			1.95& $0.74\pm0.10$&1.11 & $16.9\pm1.1$ \\
			\bottomrule[1.3pt]
		\end{tabular}	
		\label{tab:saphir}
	\end{center}
\end{table}

Due to the high luminosity of the accelerator and the large acceptance of 
the CLAS spectrometer, the CLAS data is of high precision over 
a wide $-t$ range \cite{CLAS:2013jlg,Dey:2014tfa}. 
In the following, we perform a similar analysis on the CLAS data on the proton. 
Fig. \ref{fig:phi-Clas} shows the CLAS collaboration's near-threshold
differential cross section data for $\phi$ photo-production on the hydrogen target. 
The fits to the data based on Eq. (\ref{eq:exp}) are shown in the figure. 
The differential cross section data are described reasonably well by the exponential function. 
However in the large $-t$ region, we see the cross section rising with $-t$. 
This behavior in the large $-t$ region may be due to the direct $\phi$-radiation contributions 
from $u$-channel and $s$-channel with $\phi NN$ coupling or $\phi NN^*$ coupling 
\cite{CLAS:2000kid,Laget:2000gj,Zhao:2001ue,Titov:2003bk,Titov:1998tx,Titov:1999eu,Williams:1998ge,Oh:2001bq}.

\begin{table}[htbp]
	\caption{Forward cross section $\frac{d\sigma}{dt}|_{-t=0}$ and trace anomaly $M_a/M_N$ extracted
             by fitting the experimental data from CLAS collaboration (corresponding to Fig. \ref{fig:phi-Clas}).
             Only statistical uncertainties are considered.
             The systematic uncertainty analysis is described in Sec. \ref{sec:error}.}
	\begin{center}
		\begin{tabular}{ |c|c|c|c|}
			\toprule[1.3pt]
			$E_\gamma$ (GeV) & $\frac{d\sigma}{dt}|_{-t=0}$ ($\mu$b/GeV$^2$)& $\chi^2/d.o.f.$&$M_a/M_N$ $(\%)$ \\
			\hline
			1.65& $0.93\pm0.10$&1.28&16.0$\pm$0.9\\
			\hline
			1.75& $0.89\pm0.06$&1.70 & 16.6$\pm$0.6 \\
			\hline
			1.85 &$1.18\pm0.06$ & 2.41& 20.2$\pm$0.5 \\
			\hline
			1.95 &$1.52\pm0.07$ & 3.04& 24.2$\pm$0.6 \\
			\bottomrule[1.3pt]
		\end{tabular}	
		\label{tab:clas}
	\end{center}
\end{table}

\begin{figure}[htbp]
	\centering
	\includegraphics[width=0.45\textwidth]{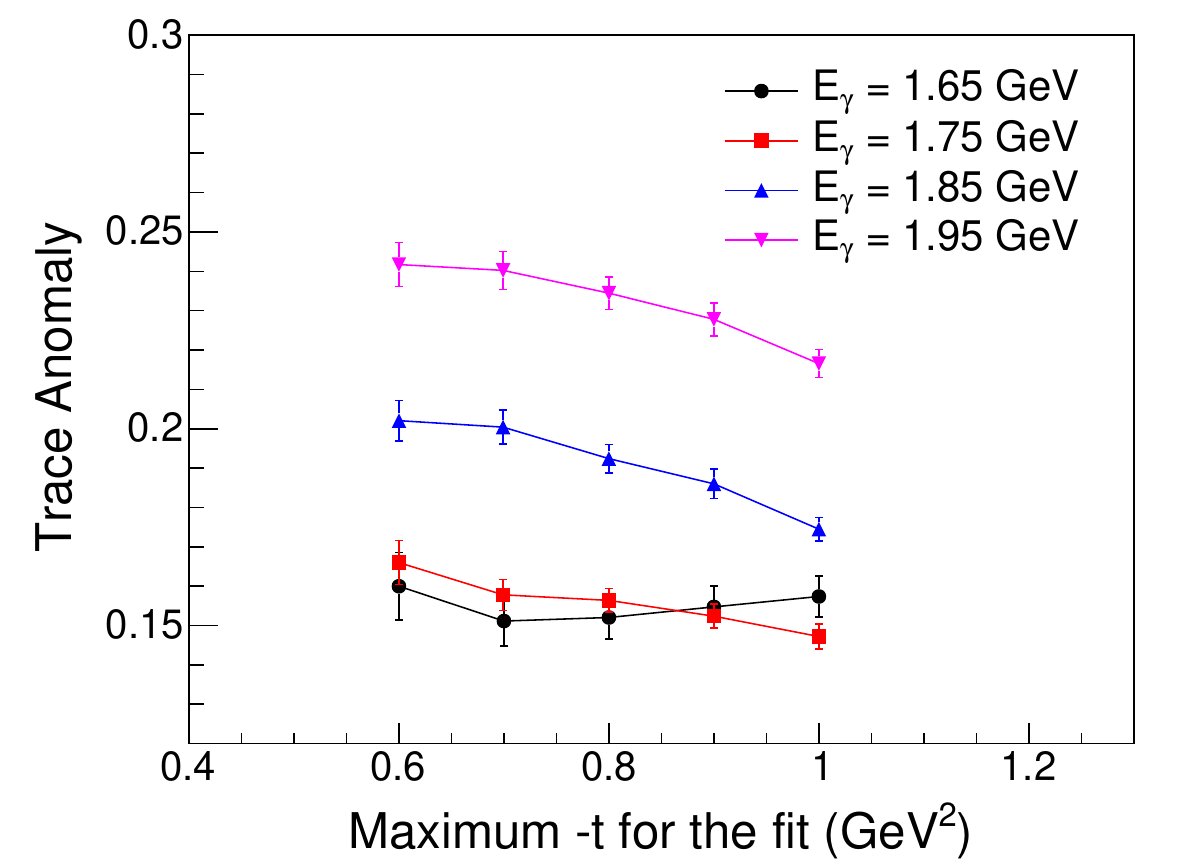}
	\caption{
		The proton trace anomaly as a function of the maximum
		$-t$ in the fitting range, extracted from CLAS data.
	}
	\label{fig:compare-t}
\end{figure}

The fitting qualities and extraction of trace anomaly are summarized in TABLE \ref{tab:clas}. 
To avoid the $u$-channel or $s$-channel contamination, 
we narrow the fit range to the small $-t$ region. 
We carefully studied the fits restricted to different $-t$ ranges, 
as the CLAS data covers a wide $-t$ range and of high precision. 
To understand the effect of large $-t$ data 
on the extraction of forward differential cross section and trace anomaly, 
we perform a series of fits excluding the large $-t$ data 
requiring $-t<0.6$ GeV$^2$, $-t<0.7$ GeV$^2$, $-t<0.8$ GeV$^2$, $-t<0.9$ GeV$^2$ and $-t<1.0$ GeV$^2$ respectively. 
The proton trace anomaly obtained from these fits are shown in FIG. \ref{fig:compare-t}. 
We find that below 1 GeV$^2$, the extracted trace anomaly does not strongly depend on 
the choice of large $-t$ cuts. 
This is probably because the $u$-channel or $s$-channel contribution 
only dominates in the large-$-t$ region where the error bars are comparatively large \cite{Laget:2000gj,Zhao:2001ue,Titov:2003bk,Titov:1998tx,Titov:1999eu,Williams:1998ge,Oh:2001bq}. 
Therefore, the large $-t$ data have little effect on the fits. 
The uncertainty arising from changing the fit range of $-t$ is 
consistent with the statistical uncertainty in the data.

\section{Uncertainty analysis}
\label{sec:error}

Now let us try to systematically estimate the uncertainty of the analysis in the previous section. 
From the method we used to extract the trace anomaly, 
the forward scattering cross sections are fitted from the data 
and the model calculating the $X-N$ elastic scattering amplitude (\ref{eq:amplitude}),
which result in the statistical and systematical uncertainties.
We start with the model and consider the uncertainty 
that may arise from the selection of the model's arguments. 
It can be seen in Eq. (\ref{eq:amplitude}) that the $X-N$ elastic scattering amplitude 
depends on the previously mentioned ``Bohr" radius $r_0$. 
In Eq. (\ref{eq:bohr}), the expression tells us that the ``Bohr" radius of vector meson 
depends on the quark mass $m_q$ and the strong coupling constant $\alpha_{\mathrm{s}}$. 
First, we take the results in Tables \ref{tab:Gluex} and \ref{tab:leps} 
as the benchmarks ($m_s = 0.486$ GeV and $m_c = 1.67$ GeV).
Then we consider the quark masses with $5\%$ fluctuations, i.e. $m_{q\pm}=m_q(1\pm5\%)$. 
The trace anomaly $\frac{1}{4}(1-b)$ obtained for different quark masses are 
shown in the following tables.

\begin{table}[H]
	\caption{Trace anomaly contributions (in percentage) to the proton mass
             with different values of strange quark mass based on LEPS data \cite{Mibe:2005er}.}
	\begin{center}
		\begin{tabular}{ |c|c|c|c|c|}
			\toprule[1.3pt]
			$E_\gamma$ (GeV) & $m_{s-}$ &$m_s$&$m_{s+}$&Mean Deviation \\
			\hline
			1.62& $15.9\pm4.2$&$18.5\pm4.9$&$21.4\pm5.6
			$&$2.75$\\
			\hline
			1.72& $14.1\pm2.2$&$16.5\pm2.5$&$19.1\pm2.9
			$&$2.5$\\
			\hline
			1.82& $17.2\pm2.4$&$20.0\pm2.8$&$23.2\pm3.2
			$&$3.0$\\
			\bottomrule[1.3pt]
		\end{tabular}	
		\label{tab:errorLeps}
	\end{center}
\end{table}

	\begin{table}[H]
		\caption{Trace anomaly contributions (in percentage) to the proton mass 
                 with different values of charm quark mass based on GlueX data \cite{Ali:2019lzf}.}
		\begin{center}
			\begin{tabular}{ |c|c|c|c|c|}
				\toprule[1.3pt]
				$E_\gamma$ (GeV) & $m_{c-}$ &$m_c$&$m_{c+}$&Mean Deviation \\
				\hline
				10.72& $16.5\pm2.9$&$19.2\pm3.3$&$22.3\pm3.9
				$&$2.9$\\
				\bottomrule[1.3pt]
			\end{tabular}	
			\label{tab:errorGluex}
		\end{center}
	\end{table}

Tables. \ref{tab:errorLeps} and \ref{tab:errorGluex} show the mean deviation 
of the trace anomaly for different quark masses for each $E_\gamma$ set. 
We easily get a relative deviation in the order of $15\%$, 
i.e. $5\%$ of the quark mass fluctuations cause $15\%$ deviation of the extracted trace anomaly. 
For the impact of the uncertainty of strong coupling $\alpha_{\mathrm{s}}$, 
we perform the same analysis as that with the quark mass.
We take the uncertainty of the coupling constant in Ref. \cite{Hoferichter:2015dsa} 
and fix the fluctuation of $\alpha_{\mathrm{s}}$ at the level of $5\%$. 
Then we find that the trace anomaly also has $15\%$ deviation with different strong couplings.

The extracted proton trace anomaly from the experiments also relies on the data fitting settings. 
We consider the statistical uncertainty in the experimental data in the previous discussions. 
As a supplement, we try to discuss some additional uncertainties that exist in terms of experiments. 
We consider the standard deviations associated with different fitting ranges 
for the Mandelstam's variable $|-t|$ for different energy cases of the CLAS data \cite{CLAS:2013jlg,Dey:2014tfa}. 
A series of fits excluding the large $-t$ data requiring $-t<0.6$ GeV$^2$, $-t<0.7$ GeV$^2$, 
$-t<0.8$ GeV$^2$, $-t<0.9$ GeV$^2$ and $-t<1.0$ GeV$^2$ are performed
and the standard deviations for different $E_\gamma$ are listed in Table.\ref{tab:stdCLAS}.

\begin{table}[htbp]
	\caption{The standard deviations of trace anomaly contributions (in percentage) 
             with different fitting ranges $|-t|<|-t_{max}|$ at different energies $E_\gamma$ 
             based on CLAS data \cite{CLAS:2013jlg,Dey:2014tfa}.}
	\begin{center}
		\begin{tabular}{ |c|c|c|c|c|}
			\toprule[1.3pt]
			$E_\gamma$ (GeV) & 1.65 &1.75&1.85&1.95\\
			\hline
			Std. Dev. & 0.37&0.70&1.13&1.03\\
			\bottomrule[1.3pt]
		\end{tabular}	
		\label{tab:stdCLAS}
	\end{center}
\end{table}

In addition, since the energy near threshold is a ambiguous definition, 
the bins between different input photon energies are relatively small, 
but there are also some variations of the result among the different data at different photon energies. 
We calculate the standard deviations from the different photon energies 
for the extraction of trace anomaly by the experimental data 
of different collaborations, which is summarised in Table. \ref{tab:stdegamma}.

\begin{table}[H]
	\caption{The standard deviations of trace anomaly contributions (in percentage) 
             with different energies $E_\gamma$ from different Collaborations 
             (CLAS: $E_\gamma=1.65$, 1.75, 1.85, 1.95 GeV; 
             SAPHIR: $E_\gamma=1.70$, 1.95 GeV; LEPS: $E_\gamma=1.62$, 1.72, 1.82 GeV).}
	\begin{center}
		\begin{tabular}{ |c|c|c|c|}
			\toprule[1.3pt]
			Collaborations & CLAS &SAPHIR&LEPS\\
			\hline
			Std. Dev. & 5.84&2.57&9.22\\
			\bottomrule[1.3pt]
		\end{tabular}	
		\label{tab:stdegamma}
	\end{center}
\end{table}

Through the demonstrations and discussions in this section, 
we find that the extraction of the proton trace anomaly from experimental data 
using the model mentioned in this work is dependent on 
the QCD parameters, especially the quark mass $m_q$ and 
the strong coupling constant $\alpha_{\mathrm{s}}$. 
Based on all the uncertainty analyses discussed above, 
we give the relative statistical uncertainties of 17.2$\%$, 17.7$\%$, 3.6$\%$, and 8.2$\%$,
the total relative systematic uncertainties of 21.4$\%$, 54.2$\%$, 37.2$\%$, and 25.7$\%$, 
for the trace anomaly determinations from the GlueX, LEPS, CLAS, and SAPHIR data, respectively.

\section{Conclusions and discussions}
\label{sec:summary}

In the previous sections we describe in detail about the VMD model 
\cite{Kharzeev:1995ij,Kharzeev:1996tw,Kharzeev:1998bz}. 
Then we propose how to obtain the trace anomaly in the proton mass decomposition 
using $\phi$ and J/$\psi$ near-threshold photo-production experimental data 
\cite{Ali:2019lzf,Mibe:2005er,Barth:2003bq,CLAS:2013jlg}. 
Based on our extracted results, we give a value of the trace anomaly inside the proton,
of which the percentage is from $16\%$ to $24\%$. 
Previous Lattice QCD study gives the trace anomaly value about $23\%$ \cite{Yang:2018nqn}. 
This work is an attempt to extract the proton trace anomaly using more types of data, 
and we find that the trace anomaly maybe not depend on the type of the vector meson probe.

The trace anomaly $M_a/M_N=0.25(1-b)$ is very sensitive to the parameter $r_0$, 
the ``Bohr" radius of vector meson. 
By Eq. (\ref{eq:bohr}) we find that it eventually depends on 
the constituent quark mass $m_q$ and the running coupling constant $\alpha_{\mathrm{s}}$. 
Thus we provide a complete uncertainty analysis in Sec. \ref{sec:error}. 
It also indicates that the trace anomaly we extracted is model parameter dependent, 
based on the VMD model adopted in this analysis. 
The VMD model is successful in describing the vector meson photo-production process, 
at least as the first step of more extensive theoretical studies. 
In the VMD model, a real photon fluctuates into a virtual vector meson, 
which subsequently scatters off the proton target. 
This model assumption is based on the fact that the vector meson has
the same quantum numbers of the photon. 
The VMD model was used for determining the $\phi-p$, $\omega-p$ and J/$\psi-p$ scattering lengths 
\cite{Strakovsky:2014wja,Strakovsky:2019bev,Strakovsky:2020uqs,Pentchev:2020kao}. 
But the applicability of the VMD model in this case requires special attentions \cite{Pentchev:2020kao}. 
For a critical review of the VMD model, the papers by Boreskov et. al. \cite{Boreskov:1976dj,Boreskov:1992ur} 
and the references therein give a very comprehensive discussions. 
In this work, the values of quark mass and running coupling are based on 
the assumptions in the non-perturbative energy region.
The strong coupling $\alpha_{\mathrm{s}}$ saturates around 3, 
which is constrained by the calculation of Lattice QCD \cite{Cui:2019dwv}. 
For the ``Bohr'' radius, more experimental constraints should be found 
in the future to achieve the goal of reducing the uncertainty. 
Higher-twist calculations should be investigated as well. 
From this work, we give the result for the proton anomaly energy,
which is similar to the result of the lattice point QCD, 
and for the first time we provide the statistical and systematical uncertainty analyses.
The model uncertainties generated by the current method in this work are significant 
and more statistics are needed on the experimental side in future.

Furthermore, the $\Upsilon$(1S) \cite{Strakovsky:2021vyk} photo-production data 
generated by Electron-Ion Colliders in the United States and China 
\cite{Accardi:2012qut,AbdulKhalek:2021gbh,Chen:2018wyz,Chen:2020ijn,Anderle:2021wcy} 
will become even more important in the future. 
Since the ``Rydberg" energy $\epsilon_0$ of $\Upsilon$(1S) is higher than 
that of other quarkonia, Eq. (\ref{eq:amplitude}) is more valid 
for the near threshold $\Upsilon$(1S) production \cite{Kharzeev:1996tw,Kharzeev:1998bz}. 
The theoretical uncertainty from the strong coupling $\alpha_{\mathrm{s}}$ is much smaller 
at a higher energy scale $\epsilon_0$. 
Therefore, the theoretical framework in this work is more suitable 
for the analysis of near-threshold  $\Upsilon$(1S) photo-production \cite{Wang:2019mza} in the future.

\begin{acknowledgments}
We are very grateful to Prof. Fan WANG for his suggestions and the discussions.
This work is supported by the Strategic Priority Research Program of Chinese Academy of Sciences
under the Grant NO. XDB34030301
and the National Natural Science Foundation of China under the Grant NO. 12005266.
\end{acknowledgments}

\bibliographystyle{apsrev4-1}
\bibliography{refs}

\end{document}